# Propagation dynamics of high-gain vortex beams in symmetry-broken media via forward and backward three-wave mixing

Fan Meng, Hao Zhu, Xin-Yao Huang†, Guo-Feng Zhang*

*School of Physics, Beihang University, Beijing 102206, China*

## Abstract

In recent years, vortex light, a distinctive form of structured light, has attracted considerable attention owing to its unique properties and the rich physical phenomena arising from its interaction with matter. In this paper, we investigate the propagation dynamics of vortex beams in a symmetry-broken three-level system based on forward and backward three-wave mixing (TWM) processes. We find that both processes enable the transfer of high-gain vortex light, with the associated topological charges obeying identical algebraic relations. The forward process exhibits periodic oscillatory transmission and modulates the transverse spatial profile of the generated signal vortex field, whereas the backward process features stable transmission and produces a signal field with higher gain and improved fidelity. Under the Autler-Townes splitting (ATS) regime, the probe field detuning in both schemes has a negligible influence on the gain. Optical depth influences only the rate at which the gain approaches its maximum, while the peak value remains unchanged. These results constitute a meaningful extension of the work reported in reference [40] and may provide a feasible approach for quantum communication, quantum computation, and the generation of high-gain, high-fidelity vortex light.

## 1. Introduction

Coherent manipulation of light-matter interactions constitutes a central topic in modern physics and has given rise to a variety of intriguing phenomena, including electromagnetically induced transparency (EIT) [1], coherent population trapping (CPT) [2], Kerr nonlinearity [3], optical solitons [4,5], four-wave mixing [6,7], all-optical switching [8], and optical bistability [9]. These advances underpin broad application prospects in areas such as quantum optics [10] and quantum information science [11]. Among these phenomena, EIT has attracted considerable attention owing to its ability to render an otherwise strongly absorbing medium transparent within a narrow spectral window [12]. EIT enables a wide range of applications, including enhanced four-wave mixing [13,14], inversion-free lasing [15], slow and fast light propagation [16,17], and refractive-index enhancement [18,19]. Corresponding theoretical and experimental studies have been carried out in a variety of platforms, including atomic systems [20], molecular magnets [21], semiconductor quantum wells [22], and optomechanical systems [23]. Although Autler-Townes splitting (ATS) exhibits transparency features similar to those of EIT, the two phenomena arise from fundamentally different physical mechanisms. The distinction between EIT and ATS has been extensively investigated and discussed in the literature [24-27]. Previous studies have generally attributed the enhancement of optical nonlinearities primarily to EIT, whereas the contribution of ATS to nonlinear optical processes has received comparatively less attention. However, M. S. Zubairy and co-workers demonstrated that ATS can play a dominant role in enhancing three-wave mixing (TWM) processes, with an efficiency substantially exceeding that achievable via EIT [28]. This finding challenges the conventional view

† Corresponding author. E-mail: xinyaohuang@buaa.edu.cn

* Corresponding author. E-mail: gf1978zhang@buaa.edu.cn

that EIT is the primary mechanism responsible for nonlinear optical enhancement.

Vortex light, an emerging form of structured light, has attracted considerable attention [29]. Its phase distribution exhibits a spiral structure due to the presence of a phase factor $e^{il\varphi}$, where $l$ denotes the topological charge, which may take integer or non-integer values, with positive and negative values corresponding to the rotation direction of the spiral wavefront, and $\varphi$ represents the azimuthal angle in the beam cross-section. Such a beam carries $l\hbar$ of orbital angular momentum (OAM), with its intensity exhibiting a dark central core corresponding to a phase singularity, where the phase is undefined. Owing to these unique properties, vortex light gives rise to numerous intriguing phenomena upon interaction with matter, including ultra-high-precision atomic positioning [30], entanglement and squeezing of OAM states [31], super-diffraction-limited effects [32], photoinduced torque [33-35], and spatially dependent EIT [36]. The transfer of vortex light OAM across different frequencies via nonlinear processes has attracted significant interest [37-41]. This process not only provides a powerful tool for quantum information transmission using OAM encoding, but also enables the generation of vortex light at frequency bands—such as far-infrared or ultraviolet—that are inaccessible via standard optical elements. While most previous studies have focused on OAM transfer in natural systems, investigations into the interaction between vortex light and symmetry-broken systems remain limited. Although certain studies have demonstrated that a three-level symmetry-broken system under the ATS effect can efficiently transfer vortex light [40], they have focused solely on the forward three-wave mixing (FTWM) process, leaving the transmission characteristics and attenuation dynamics of both forward and backward processes largely unexplored. Here, we show that the backward three-wave mixing (BTWM) process outperforms previously reported schemes in terms of gain and fidelity.

In this paper, we investigate forward and backward TWM processes under the ATS effect in a symmetry-broken three-level ladder system, yielding previously unreported results. We find that both schemes enable the transfer of high-gain vortex light OAM, with the topological charges obeying identical algebraic relations throughout the transfer. Our focus is on the transmission characteristics and attenuation dynamics of vortex light under these two schemes, revealing that: (1) In the FTWM process, vortex light transmission exhibits periodic oscillations, with the period tunable via relevant system parameters. The output intensity can therefore be modulated on demand. Moreover, an increase in the decay rate $\Gamma_{21}$ reduces the signal field intensity and alters the lateral intensity distribution through modifications of the phase structure. (2) In the BTWM process, vortex light exhibits stable transmission, contrasting sharply with the forward oscillatory behavior. Although an increase in $\Gamma_{21}$ reduces the overall signal field intensity, it does not affect the phase distribution, and consequently leaves the lateral intensity profile unchanged. Furthermore, under strong control field ATS conditions, the effect of detuning $\Delta_p$ on gain is negligible for both schemes, while the optical depth affects only the rate at which the maximum gain is reached, without altering its peak value. Compared with the forward process, the backward scheme exhibits higher signal field gain, improved fidelity, more stable transmission, and pronounced anti-loss characteristics. These findings further advance the understanding of nonlinear interactions between vortex light and symmetry-broken systems, and offer a promising approach for quantum communication, quantum computing, and the generation of high-gain, high-fidelity vortex light.

## 2. Theoretical model

As illustrated in Fig. 1, we consider a three-level ladder-type system with broken inversion symmetry, in which the weak probe field $\Omega_p$ couples the $|1\rangle \leftrightarrow |2\rangle$ transition, while the strong control field $\Omega_c$ drives the $|2\rangle \leftrightarrow |3\rangle$ transition. In a conventional three-level ladder system, the eigenstates possess well-defined parity. Owing to the SO(3) or SO(4) symmetry of the system, one-photon transitions between two energy levels are allowed only when the corresponding eigenstates have opposite parity, whereas two-photon transitions require the two states to have the same parity; these constraints are commonly referred to as dipole selection rules. Consequently, in a conventional three-level ladder system with strict spatial inversion symmetry of the potential, the one-photon transition between levels $|1\rangle$ and $|3\rangle$, which have the same parity, is forbidden, whereas the corresponding two-photon transition is allowed. However, such a two-photon process does not constitute a symmetry-broken closed-loop configuration. Therefore, a genuine three-level closed-loop system is generally difficult to realize in natural atomic or molecular energy-level structures. By contrast, when inversion symmetry of the potential is broken, the conventional dipole selection rules no longer strictly apply, thereby enabling one-photon transitions between levels $|1\rangle$ and $|3\rangle$. The breakdown of these selection rules can give rise to a variety of novel quantum phenomena, including single-photon quantum routing with dual output channels [42]. The resulting TWM process enabled by the forbidden transition offers a promising platform for investigating second-order nonlinear optical effects. Under light-matter interaction in such a three-level system, two distinct transparency mechanisms may emerge: EIT and ATS. Although both phenomena manifest as transparency windows in the absorption spectrum, they originate from fundamentally different physical mechanisms. EIT originates from Fano-type quantum interference and is fundamentally associated with destructive interference between excitation pathways, leading to the formation of a dark state. As a result, the absorption of an otherwise opaque medium is strongly suppressed, giving rise to transparency [12]. In contrast, ATS arises from the dynamic splitting of absorption resonances induced by a strong coupling field through the AC-Stark effect and does not rely on quantum interference [43]. Previous studies have largely regarded EIT as the primary mechanism responsible for enhancing optical nonlinearities, whereas the contribution of ATS has often been overlooked. However, it has been demonstrated that ATS can dominate the enhancement of TWM processes, achieving conversion efficiencies significantly exceeding those attainable under EIT conditions [28]. This finding suggests that EIT is not universally responsible for nonlinear optical enhancement. Therefore, the giant gain of the vortex beam investigated in this work is analyzed primarily within the ATS regime, where the achievable gain is substantially stronger than that in the EIT regime.

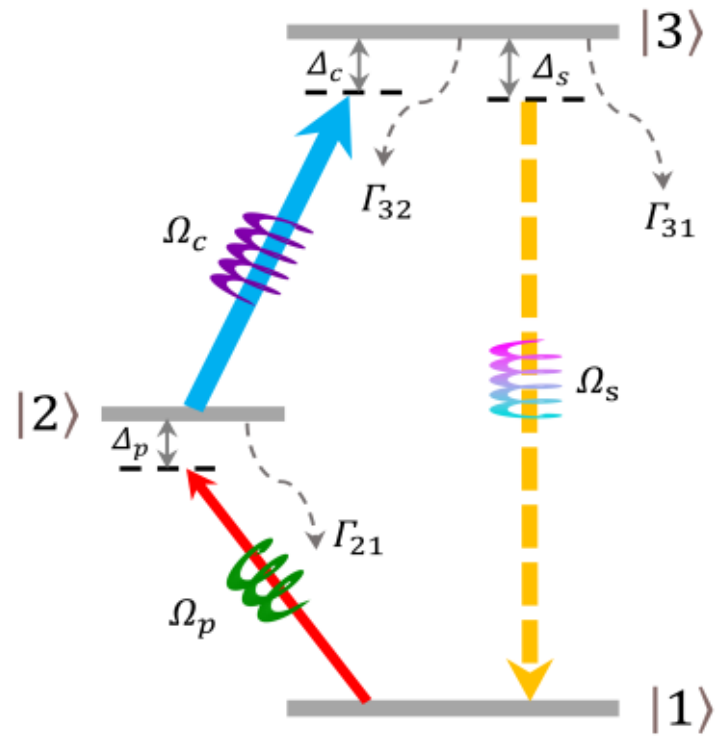

**Fig. 1.** Schematic of a symmetry-broken three-level ladder system. The weak probe field $\Omega_p$

drives the $|1\rangle \leftrightarrow |2\rangle$ transition, the strong control field $\Omega_c$ drives the $|2\rangle \leftrightarrow |3\rangle$ transition, and the generated signal field $\Omega_s$ couples the $|1\rangle \leftrightarrow |3\rangle$ transition. Here, $\Gamma_{31}$, $\Gamma_{32}$ and $\Gamma_{21}$ denote the decay rates of the respective upper energy levels, while $\Delta_p$, $\Delta_c$ and $\Delta_s$ correspond to the detunings between the respective optical fields and their associated transitions.

In the symmetry-broken three-level ladder system under investigation, and within the rotating-wave and dipole approximations, after transforming into the rotating optical-frequency frame, the system Hamiltonian can be expressed as:

$$\begin{aligned} \hat{H} &= -\hbar\Delta_p |2\rangle\langle 2| - \hbar\left(\Delta_p + \Delta_c\right)|3\rangle\langle 3| \\ &\quad -\hbar\left(\Omega_p e^{i\mathbf{k}_p\cdot\mathbf{r}} |2\rangle\langle 1| + \Omega_c e^{i\mathbf{k}_c\cdot\mathbf{r}} |3\rangle\langle 2| + \Omega_s e^{i\mathbf{k}_s\cdot\mathbf{r}} |3\rangle\langle 1| + \text{H.c.}\right), \end{aligned} \tag{1}$$

where $\Omega_p = \mu_{21}E_p/2\hbar$, $\Omega_c = \mu_{32}E_c/2\hbar$, and $\Omega_s = \mu_{31}E_s/2\hbar$ denote the Rabi frequencies of the respective optical fields. $\mu_{mn}\left(m=2,3; n=1,2; m\neq n\right)$ denotes the electric dipole matrix element, and $E_j\left(j=p,c,s\right)$ represents the amplitude of the corresponding optical electric field. The detunings of the respective fields are defined as $\Delta_p = \omega_p - \omega_{21}$ and $\Delta_c = \omega_c - \omega_{32}$. $\omega_p$ and $\omega_c$ denote the central frequencies of the incident fields, while $\omega_{21}$ and $\omega_{32}$ correspond to the transition frequencies between the respective energy levels. $\mathbf{r}$ denotes the position vector, and $\mathbf{k}_j\left(j=p,c,s\right)$ represents the wave vector of the corresponding optical field. It should be noted that the system Hamiltonian is derived within the rotating optical-frequency frame, imposing a constraint on the detunings of the three interacting fields, namely, $\Delta_s = \Delta_p + \Delta_c$. Here, we assume that the detuning of the control field is $\Delta_c = 0$.

The dynamics of the three-level ladder system can be described by the Liouville equation:

$$\frac{\partial}{\partial t}\rho = \sum_j \Gamma_j\left(L_j\rho L_j^{\dagger} - \frac{1}{2}\left\{L_j^{\dagger}L_j,\rho\right\}\right) - \frac{i}{\hbar}\left[H,\rho\right], \tag{2}$$

where the first term on the right-hand side represents the damping component, with $L_j$ denoting the Lindblad operator associated with the $j$-th dissipation channel; and the second term corresponds to the dominant dynamical process.

By substituting Eq. (1) into Eq. (2), the explicit equations of motion for the three-level ladder system are obtained as:

$$\begin{aligned}
\frac{\partial}{\partial t}\rho_{11} &= \Gamma_{31}\rho_{33} + \Gamma_{21}\rho_{22} + i\Omega_p^*\rho_{21} + i\Omega_s^*\rho_{31} - i\Omega_p\rho_{12} - i\Omega_s\rho_{13}, \\
\frac{\partial}{\partial t}\rho_{22} &= \Gamma_{32}\rho_{33} + \Gamma_{21}\rho_{22} + i\Omega_p\rho_{12} + i\Omega_c^*\rho_{32} - i\Omega_c\rho_{23} - i\Omega_p^*\rho_{21}, \\
\frac{\partial}{\partial t}\rho_{33} &= -\left(\Gamma_{31} + \Gamma_{32}\right)\rho_{33} + i\Omega_c\rho_{23} + i\Omega_s\rho_{13} - i\Omega_c^*\rho_{32} - i\Omega_s^*\rho_{31}, \\
\frac{\partial}{\partial t}\rho_{21} &= -\left(\frac{\Gamma_{21}}{2} - i\Delta_p\right)\rho_{21} + i\Omega_p\left(\rho_{11} - \rho_{22}\right) + i\Omega_c^*\rho_{31} - i\Omega_s\rho_{23}, \\
\frac{\partial}{\partial t}\rho_{31} &= -\left[\frac{\Gamma_{31} + \Gamma_{32}}{2} - i\left(\Delta_p + \Delta_c\right)\right]\rho_{31} + i\Omega_s\left(\rho_{11} - \rho_{33}\right) + i\Omega_c\rho_{21} - i\Omega_p\rho_{32},
\end{aligned}$$

$$\frac{\partial}{\partial t}\rho_{32}=-\left[\frac{\Gamma_{31}+\Gamma_{32}+\Gamma_{21}}{2}-i\Delta_c\right]\rho_{32}+i\Omega_c\left(\rho_{22}-\rho_{33}\right)+i\Omega_s\rho_{12}-i\Omega_p^*\rho_{31}. \tag{3}$$

We consider a closed three-level system satisfying population conservation, i.e., $\rho_{11}+\rho_{22}+\rho_{33}=1$. Additionally, the density matrix satisfies the Hermitian property, $\rho_{ij}=\rho_{ji}^*$. Here, $\Gamma_{31}+\Gamma_{32}=\Gamma_3$, where $\Gamma_3$ denotes the total decay rate of the upper state $|3\rangle$. The parameters $\Gamma_{21}$, $\Gamma_{31}$, and $\Gamma_{32}$ denote the decay rates from the respective upper levels to their corresponding lower levels.

In solving this set of coupled equations, we assume that the control field $\Omega_c$ is much stronger than the probe and signal fields, i.e., $\Omega_c \gg \Omega_p, \Omega_s$. Under this approximation, the control field $\Omega_c$ is treated as continuous and time- and space-independent. We further assume that the system is initially in the ground state $|1\rangle$, such that $\rho_{11}^{(0)}=1$, $\rho_{22}^{(0)}=\rho_{33}^{(0)}=0$, and $\rho_{ij}^{(0)}=0$ $(i\neq j)$. Applying perturbative decoupling to Eq. (3), $\rho_{ij}=\rho_{ij}^{(0)}+\Omega_p\rho_{ij}^{(1)}+\Omega_s\rho_{ij}^{(1)}+...$, the steady-state analytical solutions for $\rho_{21}$ and $\rho_{31}$ are obtained as:

$$\rho_{21}=\frac{2i\Gamma_3\Omega_p+4\Delta_p\Omega_p-4\Omega_s\Omega_c^*}{\left(\Gamma_{21}-2i\Delta_p\right)\left(\Gamma_3-2i\Delta_p\right)+4\left|\Omega_c\right|^2}, \tag{4}$$

$$\rho_{31}=\frac{2i\Gamma_{21}\Omega_s+4\Delta_p\Omega_s-4\Omega_p\Omega_c}{\left(\Gamma_{21}-2i\Delta_p\right)\left(\Gamma_3-2i\Delta_p\right)+4\left|\Omega_c\right|^2}, \tag{5}$$

where the detuning of the weak probe field $\Omega_p$ is denoted by $\Delta_p$, while the detuning of the strong control field $\Omega_c$ is assumed to be zero ($\Delta_c=0$).

### 2.1 Forward three-wave mixing process

For the FTWM process, under the slowly varying envelope approximation, the Maxwell-Bloch equations governing the propagation of the probe field $\Omega_p$ and the signal field $\Omega_s$ in the medium can be written as:

$$\left(\frac{1}{c}\frac{\partial}{\partial t}+\frac{\partial}{\partial z}\right)\Omega_p=i\frac{\Gamma_{21}\alpha_p}{2L}\rho_{21}, \tag{6}$$

$$\left(\frac{1}{c}\frac{\partial}{\partial t}+\frac{\partial}{\partial z}\right)\Omega_s=i\frac{\Gamma_{31}\alpha_s}{2L}\rho_{31}, \tag{7}$$

where $\alpha_j\left(j=p,s\right)$ denotes the optical depth of the medium. For simplicity, we set $\alpha_p=\alpha_s=\alpha$, and $L$ denotes the length of the medium. The phase-matching condition has been imposed, namely, $\mathbf{k}_s=\mathbf{k}_p+\mathbf{k}_c$. It should be noted that, when the medium length $L$ is sufficiently short, the diffraction term in Maxwell's equations becomes negligible and may therefore be omitted from the governing equation [40].

By substituting Eqs. (4) and (5) into Eqs. (6) and (7), respectively, and imposing the forward boundary conditions $\Omega_p\left(z=0\right)=\Omega_p\left(0\right)$ and $\Omega_s\left(z=0\right)=0$, the steady-state solutions for the probe field $\Omega_p$ and the signal field $\Omega_s$ propagating in the medium are obtained as:

$$\Omega_p(z)=\frac{\Omega_p(0)}{2X}e^{-\frac{z\alpha\left(3\Gamma_{21}\Gamma_3-2i(2\Gamma_{21}+\Gamma_3)\Delta_p+X\right)}{4LY}}\left[\left(\Gamma_{21}\Gamma_3-4i\Gamma_{21}\Delta_p+2i\Gamma_3\Delta_p\right)\left(1-e^{\frac{z\alpha X}{2LY}}\right)+X\left(1+e^{\frac{z\alpha X}{2LY}}\right)\right], \tag{8}$$

$$\Omega_s(z)=-\frac{\Omega_p(0)\Omega_c}{X}\left[2i\Gamma_3 e^{-\frac{z\alpha\left(3\Gamma_{21}\Gamma_3-2i(2\Gamma_{21}+\Gamma_3)\Delta_p+X\right)}{4LY}}\left(-1+e^{\frac{z\alpha X}{2LY}}\right)\right], \tag{9}$$

where $X=\sqrt{\left[\Gamma_{21}\left(\Gamma_3-4i\Delta_p\right)+2i\Gamma_3\Delta_p\right]^2-32\Gamma_{21}\Gamma_3|\Omega_c|^2}$, $Y=\left(\Gamma_{21}-2i\Delta_p\right)\left(\Gamma_3-2i\Delta_p\right)+4|\Omega_c|^2$.

## 2.2 Backward three-wave mixing process

For the BTWM process, the primary differences arise from the boundary conditions and the field-propagation equations. In this configuration, the signal field counter-propagates with respect to the probe field. The propagation of the probe field $\Omega_p$ in the medium is still governed by Eq. (6), whereas the Maxwell-Bloch equation describing the evolution of the signal field $\Omega_s$ takes the form

$$\left(\frac{1}{c}\frac{\partial}{\partial t}-\frac{\partial}{\partial z}\right)\Omega_s=i\frac{\Gamma_{31}\alpha_s}{2L}\rho_{31}, \tag{10}$$

the definitions and values of the relevant parameters are the same as those used in the FTWM process. The phase-matching condition $\mathbf{k}_s=\mathbf{k}_p+\mathbf{k}_c$ remains necessary in this case.

By imposing the backward boundary conditions $\Omega_p(z=0)=\Omega_p(0)$ and $\Omega_s(z=L)=0$, the steady-state solutions for the probe field $\Omega_p$ and the signal field $\Omega_s$ propagating in the medium are obtained as:

$$\Omega_p(z)=\frac{2\Omega_p(0)e^{-\frac{z\alpha\left(2i\Gamma_3\Delta p+\Gamma_{21}(\Gamma_3-4i\Delta p)\right)}{2LA}}\left(4\Gamma_{21}\Gamma_3|\Omega_c|^2\left(e^{\frac{(2L-z)\alpha B}{2LA}}+e^{\frac{z\alpha B}{2LA}}\right)+\frac{1}{8}C\left(e^{\frac{(2L-z)\alpha B}{2LA}}(C+B)+e^{\frac{z\alpha B}{2LA}}(C-B)\right)\right)}{\frac{1}{4}B\left(\left(-1+e^{\frac{\alpha B}{A}}\right)C+\left(1+e^{\frac{\alpha B}{A}}\right)B\right)}, \tag{11}$$

$$\Omega_s(z)=\frac{4\Omega_p(0)\Omega_c\Gamma_3 e^{-\frac{z\alpha\left(2i\Gamma_3\Delta p+\Gamma_{21}(\Gamma_3-4i\Delta p)\right)}{2LA}}\left(e^{\frac{(2L-z)\alpha B}{2LA}}-e^{\frac{z\alpha B}{2LA}}\right)}{i\left(\left(-1+e^{\frac{\alpha B}{A}}\right)C+\left(1+e^{\frac{\alpha B}{A}}\right)B\right)}, \tag{12}$$

where $A=2\left(\Gamma_{21}-2i\Delta p\right)\left(\Gamma_3-2i\Delta p\right)+8|\Omega_c|^2$, $B=\sqrt{\left(-3\Gamma_{21}\Gamma_3+2i\Gamma_3\Delta_p+4i\Gamma_{21}\Delta_p{}^2\right)+32\Gamma_{21}\Gamma_3|\Omega_c|^2}$ and $C=3\Gamma_{21}\Gamma_3-2i\Gamma_3\Delta_p-4i\Gamma_{21}\Delta_p$.

Before analyzing the theoretical results of forward and backward TWM, we first introduce the Laguerre-Gaussian (LG) vortex beam, which is most commonly employed in this context. The complex amplitude describing the spatial distribution of the LG field in cylindrical coordinates can be written as [38,44]:

$$\Omega(r,\varphi)=\Omega_0\frac{1}{\sqrt{|l|!}}\left(\frac{\sqrt{2}r}{w_0}\right)^{|l|}L_{\tilde{p}}^{|l|}\left(\frac{2r^2}{w_0^2}\right)e^{-\frac{r^2}{w_0^2}}e^{il\varphi}, \tag{13}$$

where $\Omega_0$, $r$, $w_0$, $l$, and $\tilde{p}$ denote the field amplitude, radial coordinate, beam waist, topological charge, and radial index, respectively. $L_{\tilde{p}}^{|l|}$ is the associated Laguerre polynomial, which can be written as:

$$L_{\tilde{p}}^{|l|}(x)=\frac{e^x x^{-|l|}}{\tilde{p}!}\frac{d^{\tilde{p}}}{dx^{\tilde{p}}}\left(x^{|l|+\tilde{p}}e^{-x}\right), \tag{14}$$

where $x=2r^2/w_0^2$ characterizes the radial dependence of the LG beam for different radial indices. When $l\neq 0$, the LG beam carries OAM along the optical axis.

## 3. Results and Discussion

### 3.1 Transmission characteristics of vortex light

In a three-level symmetry-broken system, the influence of vortex light transmission characteristics and the system's attenuation on gain has not been fully explored in previous studies. Based on the theoretical framework presented in Section 2, we investigate in detail the vortex light transmission characteristics and the control of the attenuation, considering separately the forward and backward TWM processes. Previous studies have shown that, in symmetry-broken systems, the ATS effect produces a larger gain than the EIT effect [28], and that a smooth transition between the two can be achieved by tuning the strength of the control field $\Omega_c$. Therefore, in this work, we focus primarily on the high-gain behavior of the system under the ATS effect, with the control field set to $\Omega_c=10\Gamma_3$($|\Omega_c|>\frac{1}{2}|\Gamma_3-\Gamma_{21}|$, which falls within the ATS regime).

#### 3.1.1. Forward three-wave mixing process

In the FTWM process, we first examine the influence of the detuning $\Delta_p$ on the transmission behavior of the probe field $\Omega_p$ and the signal field $\Omega_s$ in the medium under varying decay rates $\Gamma_{21}$. Figs. 2(a)-(c) illustrate how the relative intensities $|\Omega_{p,s}(z)/\Omega_p(0)|^2$ of the probe field $\Omega_p$ and the signal field $\Omega_s$ vary with the dimensionless length $z/L$ in the medium under different values of detuning $\Delta_p$ for decay rates $\Gamma_{21}=0.25\Gamma_3$, $0.5\Gamma_3$ and $1\Gamma_3$, respectively. Specifically, Fig. 2(a) shows the propagation behavior of the probe and signal fields in the medium when $\Gamma_{21}=0.25\Gamma_3$. Both fields exhibit periodic oscillations during propagation. When the decay rate is small ($\Gamma_{21}=0.25\Gamma_3$), the generated signal field exhibits high gain, which can even exceed unity. This high gain arises because the strong control field $\Omega_c$ pumps population from energy level $|2\rangle$ to level $|3\rangle$, providing additional energy. Consequently, the assertion that the conventional definition of efficiency, as reported in previous studies [28,40], is invalid under these conditions is unwarranted. Accordingly, $|\Omega_{p,s}(z)/\Omega_p(0)|^2$ should be defined as the gain (or relative intensity), while the proper definition of efficiency should refer to Ref. [45]. Fig. 2(b) depicts the transmission characteristics of the probe and signal fields in the medium for $\Gamma_{21} = 0.5\Gamma_3$. The fields still exhibit periodic oscillations, but with a reduced oscillation period. As $\Gamma_{21}$ increases, the gain of the generated signal field decreases to below unity. This reduction originates from the reduced absorption of the probe field at larger $\Gamma_{21}$, which consequently weakens the gain associated with the $|3\rangle \leftrightarrow |1\rangle$ transition. As shown in Fig. 2(c), when the decay rate $\Gamma_{21}$ is further increased (i.e., $\Gamma_{21} = 1\Gamma_3$), the oscillation periods of both the probe and signal fields become shorter during propagation through the medium. Simultaneously, the increased decay rate $\Gamma_{21}$ significantly suppresses the amplification of the generated signal field, resulting in a further reduction in its gain. Furthermore, Figs. 2(a)–(c) show that variations in the probe detuning $\Delta_p$ have a negligible effect on the oscillation periods of both the probe and signal fields during propagation. Meanwhile, increasing the detuning $\Delta_p$ leads to a reduction in the gain of the generated signal field $\Omega_s$. However, under the strong control field $\Omega_c = 10\Gamma_3$ based on the ATS effect, this reduction is not pronounced. In all three figures, the optical depth is set to $\alpha = 300$.

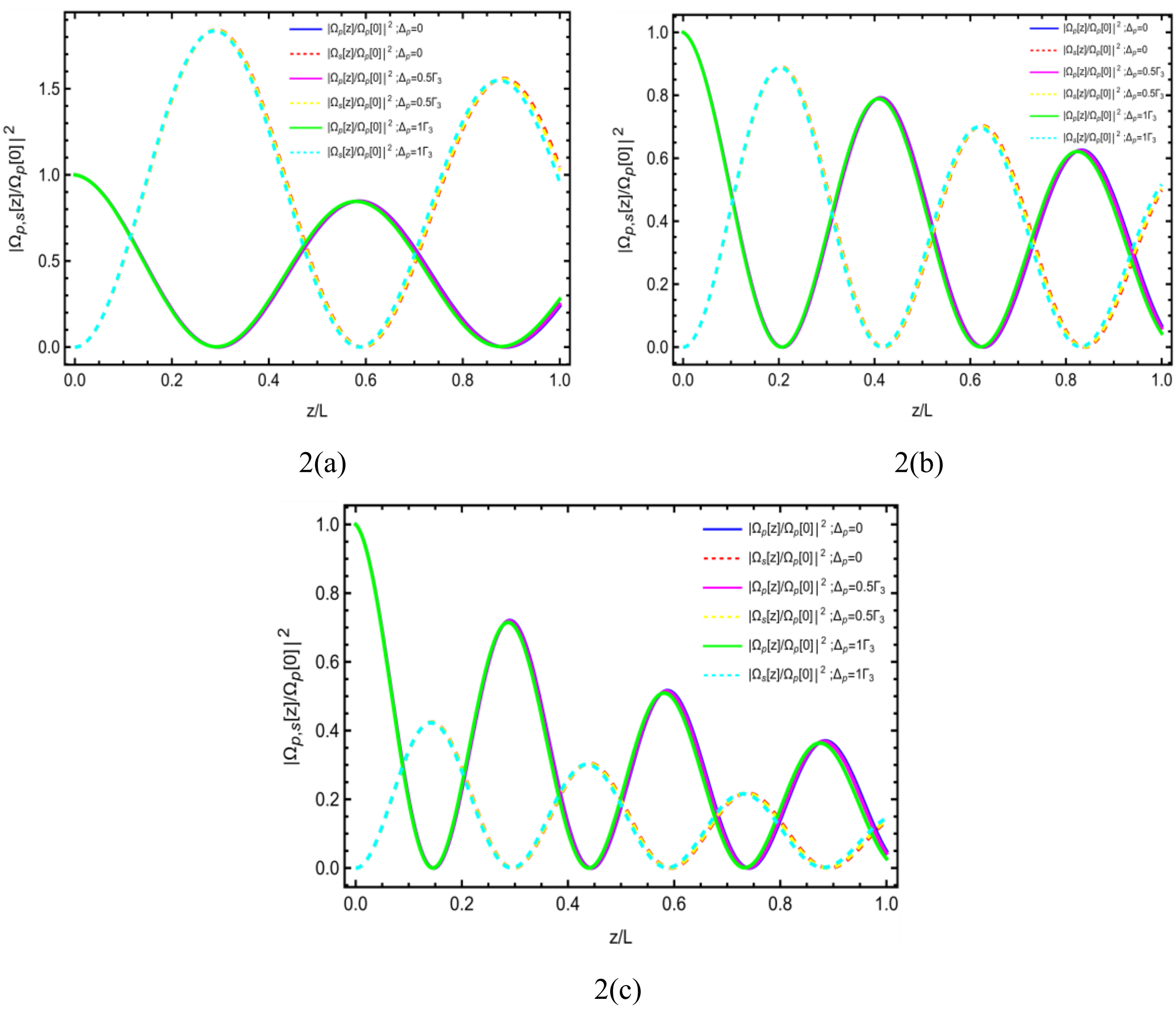

**Fig. 2.** During forward propagation, the relative intensities $\left|\Omega_{p,s}(z)/\Omega_p(0)\right|^2$ of the probe field $\Omega_p$ and the signal field $\Omega_s$ vary with the dimensionless length $z/L$ under different detuning values $\Delta_p\left(\Delta_p = 0, 0.5, 1\right)$. The effects of varying decay rates $\Gamma_{21}$ are also considered. (a) $\Gamma_{21} = 0.25\Gamma_3$, (b) $\Gamma_{21} = 0.5\Gamma_3$, (c) $\Gamma_{21} = 1\Gamma_3$. Other parameters are set as $\Omega_{p0} = 0.1\Gamma_3$, $\Omega_{c0} = 10\Gamma_3$, and $\alpha = 300$.

Next, we investigated the influence of optical depth $\alpha$ on the transmission behavior of the probe field $\Omega_p$ and the signal field $\Omega_s$ in the medium under varying decay rates $\Gamma_{21}$. Figs. 3(a)-(c) illustrate how the relative intensities $\left|\Omega_{p,s}(z)/\Omega_p(0)\right|^2$ of the probe and signal fields vary with the dimensionless length $z/L$ in the medium under different optical depths $\alpha$, for decay rates $\Gamma_{21}=0.25\Gamma_3$, $0.5\Gamma_3$ and $1\Gamma_3$. Specifically, Fig. 3(a) illustrates the propagation dynamics of the probe field and the generated signal field in the medium for a decay rate of $\Gamma_{21}=0.25\Gamma_3$. Both the probe and signal fields exhibit periodic oscillatory propagation through the medium. For a relatively small decay rate $\Gamma_{21}$, the generated signal field experiences substantial amplification, with its gain exceeding unity. As shown in Fig. 3(b), increasing the decay rate to $\Gamma_{21}=0.5\Gamma_3$ shortens the oscillation period of both fields and significantly reduces the gain of the generated signal field to below unity. As shown in Fig. 3(c), a further increase in the decay rate to $\Gamma_{21}=1\Gamma_3$ leads to an additional reduction in the oscillation period and strongly suppresses the amplification of the generated signal field, yielding a gain far below unity. Furthermore, Figs. 3(a)-(c) demonstrate that, in contrast to the negligible influence of the probe detuning $\Delta_p$, the optical depth $\alpha$ has a pronounced effect on the oscillation period of the propagating fields. Specifically, the oscillation period decreases progressively with increasing optical depth. Notably, the optical depth only affects the rate at which the maximum gain is reached, without altering the final peak value. In all three figures, the detuning is set to $\Delta_p=0$.

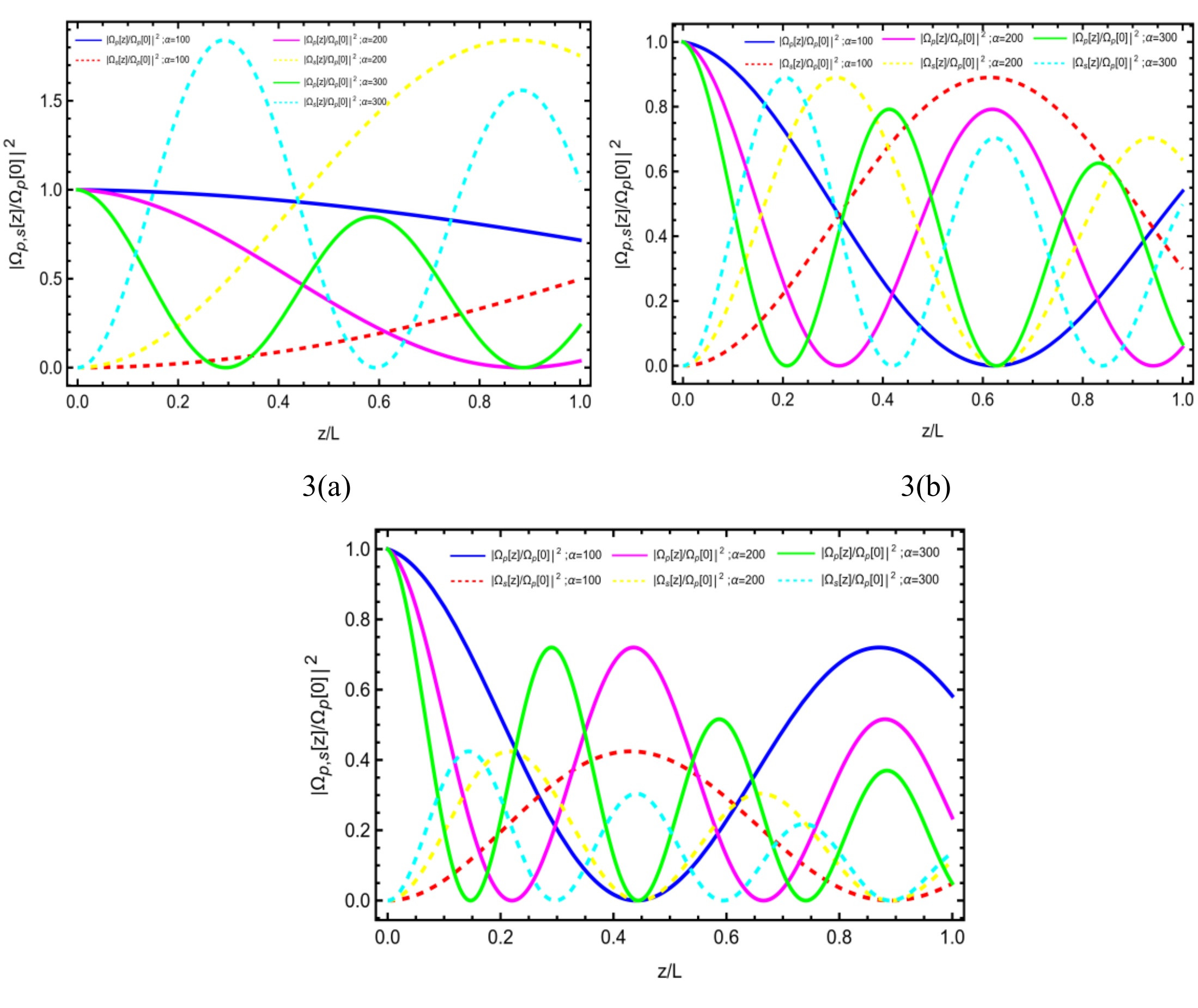

3(a) 3(b)

3(c)

**Fig. 3.** During forward propagation, the relative intensities $\left|\Omega_{p,s}(z)/\Omega_p(0)\right|^2$ of the probe field $\Omega_p$ and the signal field $\Omega_s$ vary with the dimensionless length $z/L$ under different optical depths $\alpha$ ($\alpha = 100, 200, 300$). The effects of varying decay rates $\Gamma_{21}$ are also considered. (a) $\Gamma_{21} = 0.25\Gamma_3$, (b) $\Gamma_{21} = 0.5\Gamma_3$, (c) $\Gamma_{21} = 1\Gamma_3$. Other parameters are set as $\Omega_{p0} = 0.1\Gamma_3$, $\Omega_{c0} = 10\Gamma_3$, and $\Delta_p = 0$.

The above analysis of the FTWM process demonstrates that the oscillation period of the propagating fields can be effectively controlled by tuning the decay rate $\Gamma_{21}$ and the optical depth $\alpha$. Consequently, the gain of the generated signal field at the output of the medium can be selectively accessed and optimized as required.

### 3.1.2. Backward three-wave mixing process

In the BTWM process, we first examined the influence of detuning $\Delta_p$ on the transmission behavior of the probe field $\Omega_p$ and the signal field $\Omega_s$ in the medium under varying decay rates $\Gamma_{21}$. Figs. 4(a)-(c) illustrate the evolution of the relative intensities $\left|\Omega_{p,s}(z)/\Omega_p(0)\right|^2$ of the probe and signal fields as a function of the dimensionless length $z/L$ in the medium under different detuning values $\Delta_p$ for decay rates $\Gamma_{21} = 0.25\Gamma_3$, $0.5\Gamma_3$ and $1\Gamma_3$. Specifically, Fig. 4(a) presents the evolution of the relative intensities of the probe field and the generated signal field in the medium for a decay rate of $\Gamma_{21} = 0.25\Gamma_3$. As the propagation distance increases toward the medium output ( $z \to L$ ), the relative intensity of the probe field decreases monotonically, whereas the generated signal field is progressively amplified along the propagation direction ( $z \to 0$ ). The signal-field gain can substantially exceed unity. Moreover, the gain rate of the signal field exceeds the attenuation rate of the probe field, as evidenced by the larger slope of the corresponding intensity curve. Fig. 4(b) depicts the corresponding evolution for $\Gamma_{21} = 0.5\Gamma_3$. The probe-field intensity continues to decrease while the signal-field intensity increases. In this case, the attenuation rate of the probe field is approximately equal to the gain rate of the signal field, as indicated by the nearly overlapping slopes of the two curves. Although the signal-field gain decreases with increasing $\Gamma_{21}$, it remains close to unity. This result indicates that, under identical conditions, the backward TWM process provides stronger gain than its forward counterpart. As shown in Fig. 4(c), when the decay rate is further increased to $\Gamma_{21} = 1\Gamma_3$, the probe and signal fields retain the same monotonic evolution observed in Figs. 4(a) and 4(b). However, the gain of the generated signal field is significantly reduced to a value well below unity. Under this condition, the attenuation rate of the probe field exceeds the gain rate of the signal field. Furthermore, Figs. 4(a)-(c) show that variations in the probe detuning $\Delta_p$ have only a minor influence on the evolution rates of the probe and signal field intensities. Moreover, under a strong control field $\Omega_c$ ($\Omega_c = 10\Gamma_3$), the ATS suppresses the influence of detuning, rendering its effect on the signal-field gain nearly negligible. In all three figures, the optical depth is set to $\alpha = 300$.

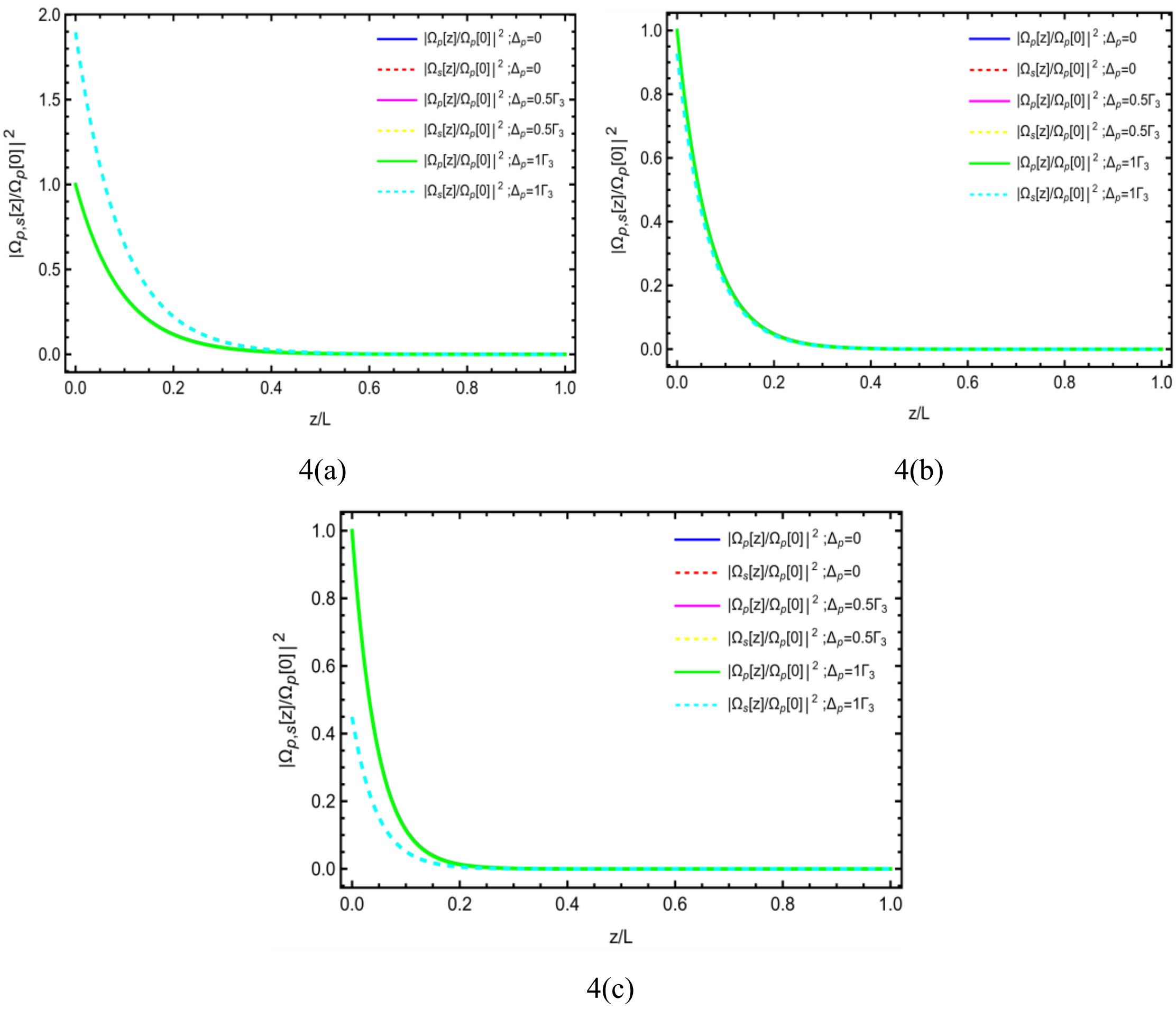

**Fig. 4.** During backward propagation, the relative intensities $\left|\Omega_{p,s}(z)/\Omega_{p}(0)\right|^{2}$ of the probe field $\Omega_p$ and the signal field $\Omega_s$ vary with the dimensionless length $z/L$ under different detuning values $\Delta_p\left(\Delta_p=0,0.5,1\right)$. The effects of varying decay rates $\Gamma_{21}$ are also considered. (a) $\Gamma_{21}=0.25\Gamma_3$, (b) $\Gamma_{21}=0.5\Gamma_3$, (c) $\Gamma_{21}=1\Gamma_3$. Other parameters are set as $\Omega_{p0}=0.1\Gamma_3$, $\Omega_{c0}=10\Gamma_3$, and $\alpha=300$.

Figs. 5(a)-(c) present the evolution of the normalized intensities of the probe field $\Omega_p$ and generated signal field $\Omega_s$ as functions of the normalized propagation distance $z/L$ for different optical depth $\alpha$, with $\Gamma_{21}=0.25\Gamma_3$, $\Gamma_{21}=0.5\Gamma_3$, and $\Gamma_{21}=0.5\Gamma_3$, respectively. Fig. 5(a) illustrates the propagation dynamics of the probe and generated signal fields for $\Gamma_{21}=0.25\Gamma_3$. As propagation proceeds, the probe-field intensity decreases monotonically, whereas the generated signal field undergoes continuous amplification. The signal-field gain substantially exceeds unity. In this regime, the gain rate of the signal field exceeds the attenuation rate of the probe field. As shown in Fig. 5(b), increasing the decay rate to $\Gamma_{21}=0.5\Gamma_3$ reduces the signal-field gain, but it remains close to unity. In this case, the attenuation rate of the probe field is nearly equal to the growth rate of the signal field. When the decay rate is further increased to $\Gamma_{21}=1\Gamma_3$, the signal-field gain decreases markedly and falls below unity, as shown in Fig. 5(c). In this regime, the increased decay rate strongly suppresses the energy-transfer process driven by the control field $\Omega_c$. Consequently, the attenuation rate of the probe field exceeds the gain rate of the signal field, and the overall dynamics become attenuation-dominated. Furthermore, Figs. 5(a)-(c) indicate that, in contrast to the probe detuning $\Delta_p$, the optical depth $\alpha$ strongly influences the evolution rate of the field intensities. Specifically, increasing the optical depth accelerates the spatial evolution of both fields. However, the optical depth affects only the rate at which the gain approaches its maximum value, while leaving the ultimate peak gain essentially unchanged. In all three figures, the detuning is set to $\Delta_p = 0$.

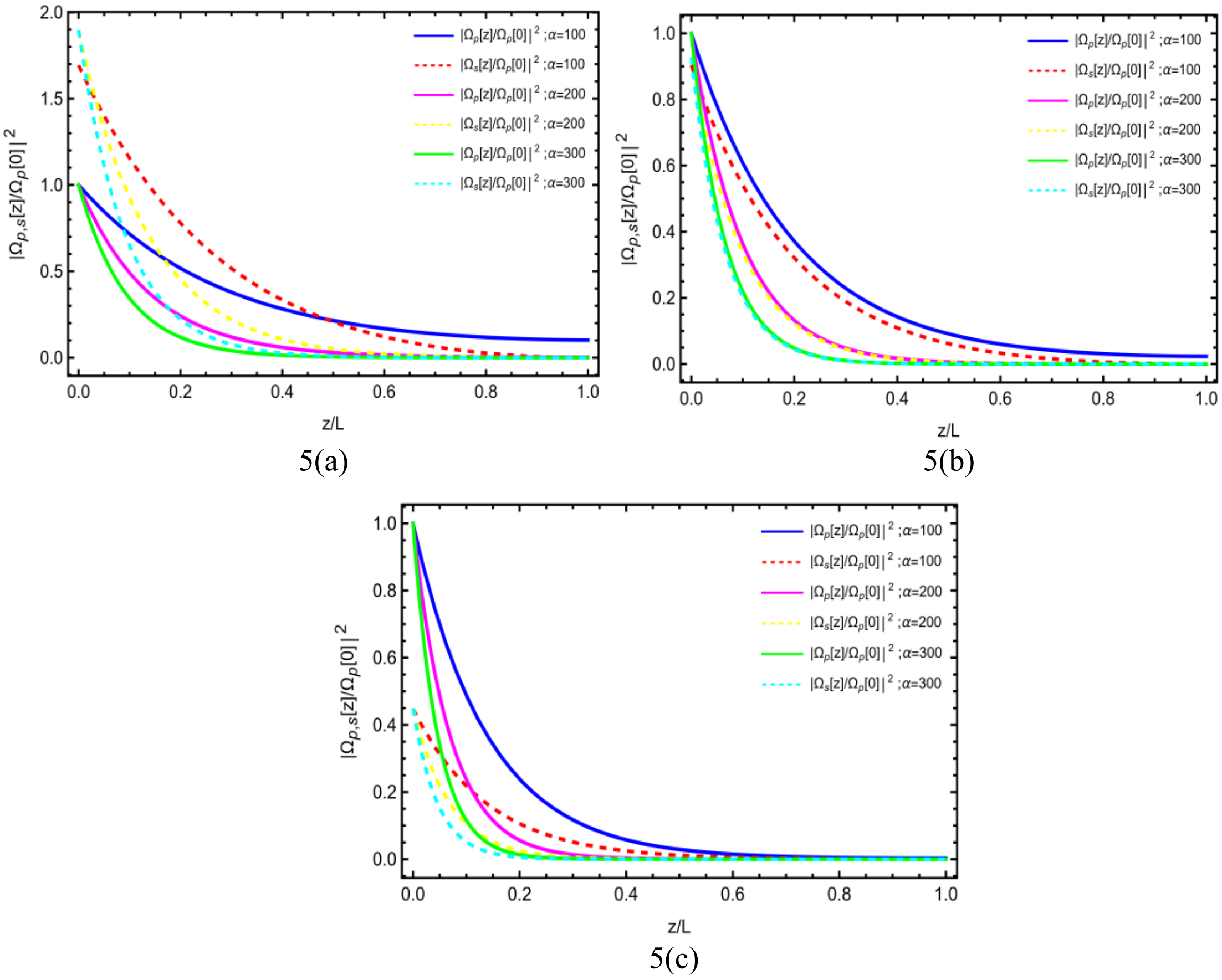

**Fig. 5.** During backward propagation, the relative intensities $\left|\Omega_{p,s}(z)/\Omega_p(0)\right|^2$ of the probe field $\Omega_p$ and the signal field $\Omega_s$ vary with the dimensionless length $z/L$ under different optical depths $\alpha$ $(\alpha = 100, 200, 300)$. The effects of varying decay rates $\Gamma_{21}$ are also considered. (a) $\Gamma_{21} = 0.25\Gamma_3$, (b) $\Gamma_{21} = 0.5\Gamma_3$, (c) $\Gamma_{21} = 1\Gamma_3$. Other parameters are set as $\Omega_{p0} = 0.1\Gamma_3$, $\Omega_{c0} = 10\Gamma_3$, and $\Delta_p = 0$.

The above analysis of the BTWM process reveals propagation dynamics that are fundamentally different from the periodic oscillatory behavior observed in the forward configuration. Specifically, the probe-field intensity decreases monotonically during propagation, whereas the signal-field intensity increases continuously, resulting in a stable and monotonic energy-transfer process. Moreover, under identical operating conditions, the BTWM process provides substantially stronger gain of the generated signal field than its forward counterpart.

### 3.2. Transfer of vortex light

According to the theoretical results presented in Section 2, the FTWM process is described by Eqs. (8) and (9), while the BTWM process is governed by Eqs. (11) and (12). It is evident that both the forward and backward TWM processes enable the transfer of OAM carried by vortex light. In the following, we analyze the OAM transfer characteristics in these two configurations and investigate the influence of the decay rate $\Gamma_{21}$ on the transfer process.

### 3.2.1. Forward three-wave mixing process

We first consider the case in which only one of the two input fields carries a vortex phase. Without loss of generality, we assume that only the probe field $\Omega_p$ is a vortex beam. According to Eq. (9), when the initial probe field $\Omega_p(0)$ takes the form given in Eq. (13), the generated signal field $\Omega_s$ also exhibits a vortex structure, satisfying $\Omega_s \sim \Omega_p(0)$. This indicates that the OAM of the vortex beam can be transferred from the incident probe field to the generated signal field. Fig. 6(a) illustrates the evolution of the intensity and phase distributions of the generated signal field as a function of the decay rate $\Gamma_{21}$, when the probe field is prepared in the LG mode $LG_2^3(\tilde{p}_p = 2, l_p = 3)$ in the forward configuration. Specifically, as $\Gamma_{21}$ increases, the intensity of the generated signal field decreases, with the outer ring experiencing pronounced attenuation. This behavior arises from the progressive fragmentation of the vortex phase into multiple regions with increasing $\Gamma_{21}$, which in turn modifies the transverse intensity distribution. As can be seen, although the phase is divided into multiple regions, a total phase winding of $2\pi l_s$ is preserved around the phase singularity, clearly demonstrating the topological robustness of vortex beams against attenuation [46].

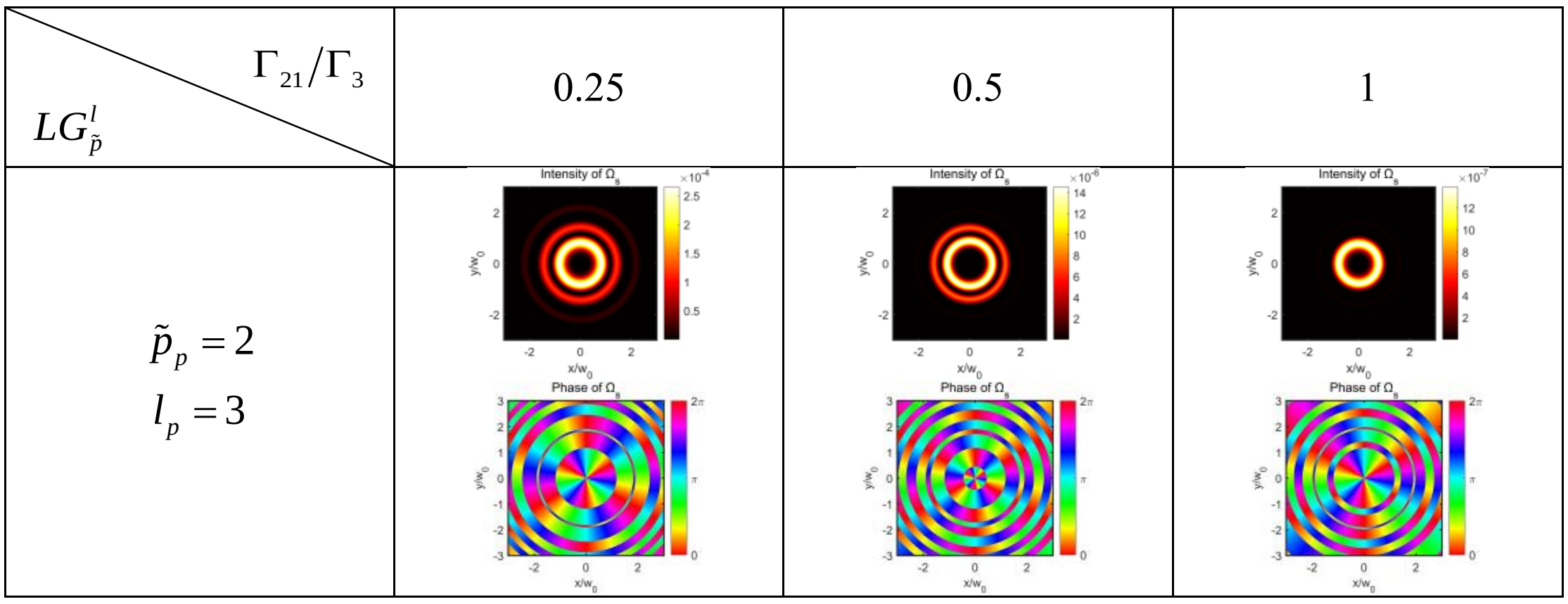

**Fig. 6(a).** During forward propagation, when the probe field $\Omega_p$ is prepared in a LG vortex mode with $LG_2^3(\tilde{p}_p = 2, l_p = 3)$ and the control field $\Omega_c$ is a fundamental Gaussian beam, the intensity and phase distributions of the generated signal field $\Omega_s$ are shown for different decay rates $\Gamma_{21}$. Other parameters are set as $\Omega_{p0} = 0.1\Gamma_3$, $\Omega_{c0} = 10\Gamma_3$, $w_{p0} = w_0$, $w_{c0} = 3w_0$, $\Delta_p = 0$, $\alpha = 100$.

Next, we consider the case where both the input probe field $\Omega_p$ and the control field $\Omega_c$ are vortex beams. According to Eq. (9), when $\Omega_p(0)$ and $\Omega_c$ both take the form of Eq. (13), the generated signal field $\Omega_s$ also exhibits a vortex structure, satisfying $\Omega_s \sim \Omega_p(0)\Omega_c$. Consequently, the OAM carried by the signal field originates from both input fields, and the corresponding topological charge obeys the conservation relation $l_s = l_p + l_c$. Fig. 6(b) illustrates the evolution of the intensity and phase distributions of the generated signal field as the decay rate $\Gamma_{21}$ increases, where the probe field is prepared in a LG mode $LG_0^3\left(\tilde{p}_p=0,l_p=3\right)$ and the control field in $LG_0^2\left(\tilde{p}_c=0,l_c=2\right)$. As $\Gamma_{21}$ increases, the intensity of the vortex beam diminishes progressively across the transverse profile. Meanwhile, the phase distribution is segmented into multiple regions, leading to corresponding modifications in the transverse intensity distribution. Nevertheless, the $2\pi l_s$ phase winding around the phase singularity remains well preserved, clearly demonstrating the robustness of OAM transfer against attenuation. For simplicity, the influence of the radial index $p$ of the vortex modes is neglected in this analysis.

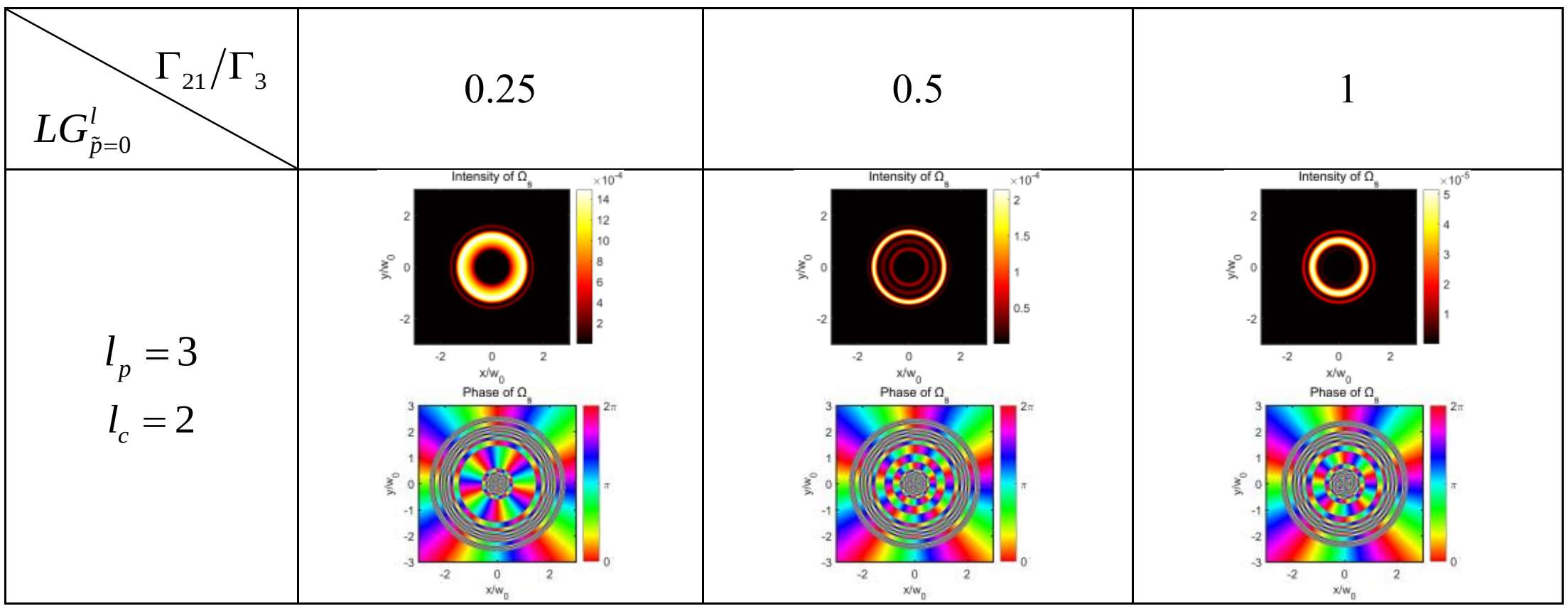

**Fig. 6(b).** During the forward propagation process, when the probe field is prepared in a LG mode $LG_0^3\left(\tilde{p}_p=0,l_p=3\right)$ and the control field in $LG_0^2\left(\tilde{p}_c=0,l_c=2\right)$, the intensity and phase distributions of the generated signal field as functions of the decay rate $\Gamma_{21}$ are shown. Other parameters are set as $\Omega_{p0}=0.1\Gamma_3$, $\Omega_{c0}=10\Gamma_3$, $w_{p0}=w_0$, $w_{c0}=w_0$, $\Delta_p=0$, $\alpha=100$.

### 3.2.2. Backward three-wave mixing process

We first consider the case in which only the probe field $\Omega_p$ carries vortex structure. According to Eq. (12), the generated signal field $\Omega_s$ inherits the same vortex characteristics as the probe field ($\Omega_s\sim\Omega_p(0)$), indicating faithful transfer of OAM from $\Omega_p$ to $\Omega_s$. Fig. 7(a) illustrates the evolution of the intensity and phase distributions of the generated signal field as a function of the decay rate $\Gamma_{21}$, when the probe field is prepared in the vortex mode $LG_2^3\left(\tilde{p}_p=2,l_p=3\right)$ during the backward process. Specifically, as $\Gamma_{21}$ increases, the intensity of the generated signal field decreases, with the attenuation less pronounced than in the forward process under identical conditions. Moreover, in contrast to forward propagation, variations in $\Gamma_{21}$ do not modify the phase distribution in the backward process, such that the transverse intensity profile is well preserved. These results demonstrate that the BTWM process is more effective for vortex light transfer and can substantially enhance the transfer fidelity.

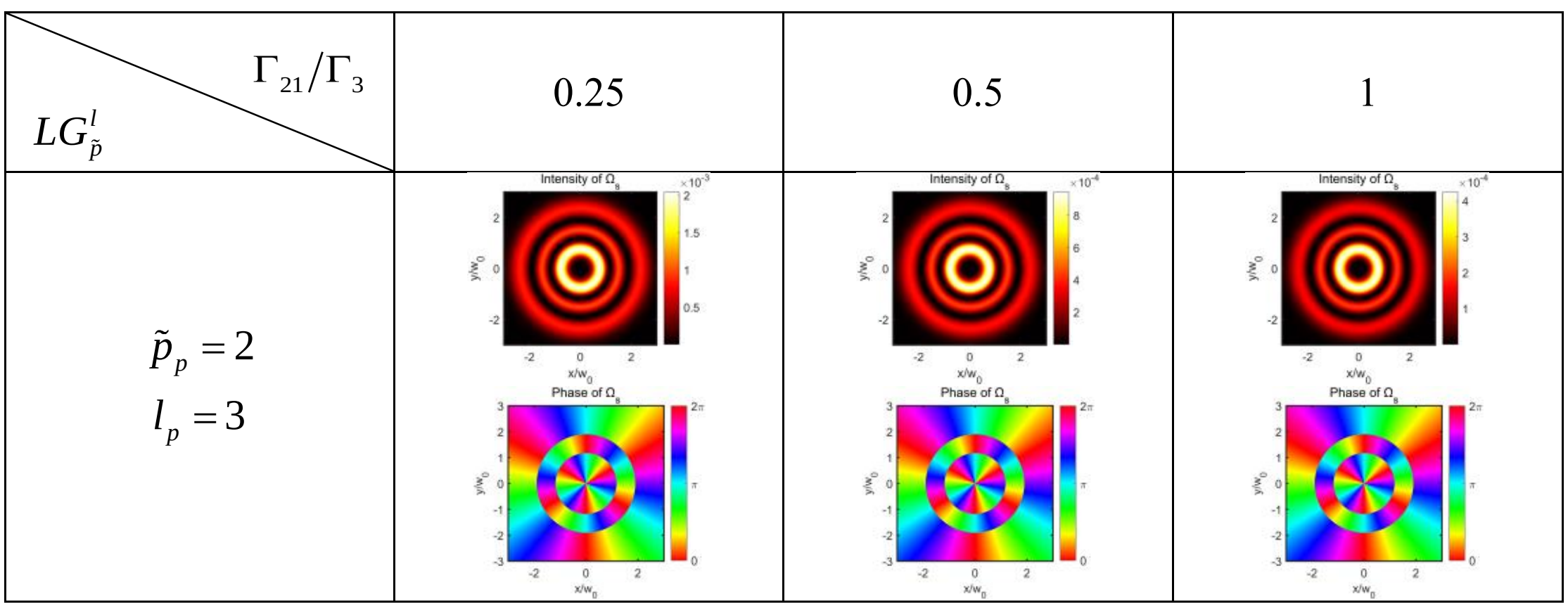

**Fig. 7(a).** During backward propagation, when the probe field $\Omega_p$ is prepared in a LG vortex mode with $LG_2^3(\tilde{p}_p = 2, l_p = 3)$ and the control field $\Omega_c$ is a fundamental Gaussian beam, the intensity and phase distributions of the generated signal field $\Omega_s$ are shown for different decay rates $\Gamma_{21}$. Other parameters are set as $\Omega_{p0} = 0.1\Gamma_3$, $\Omega_{c0} = 10\Gamma_3$, $w_{p0} = w_0$, $w_{c0} = 3w_0$, $\Delta_p = 0$, $\alpha = 100$.

Next, we consider the situation in which both the probe field and the control field carry vortex structures. According to Eq. (12), the generated signal field $\Omega_s$ accumulates the OAM of the probe and control fields ($\Omega_s \sim \Omega_p(0)\Omega_c$), such that the topological charge satisfies $l_s = l_p + l_c$. Fig. 7(b) illustrates the evolution of the intensity and phase distributions of the generated signal field as a function of the decay rate $\Gamma_{21}$, when the probe field is prepared in the vortex mode $LG_0^3(\tilde{p}_p = 0, l_p = 3)$ and the control field in $LG_0^2(\tilde{p}_c = 0, l_c = 2)$. Specifically, as $\Gamma_{21}$ increases, the intensity distribution of the vortex signal field gradually diminishes, while remaining less attenuated than in the forward process. Moreover, the phase distribution remains insensitive to variations in $\Gamma_{21}$, thereby substantially improving the fidelity of the transverse intensity profile.

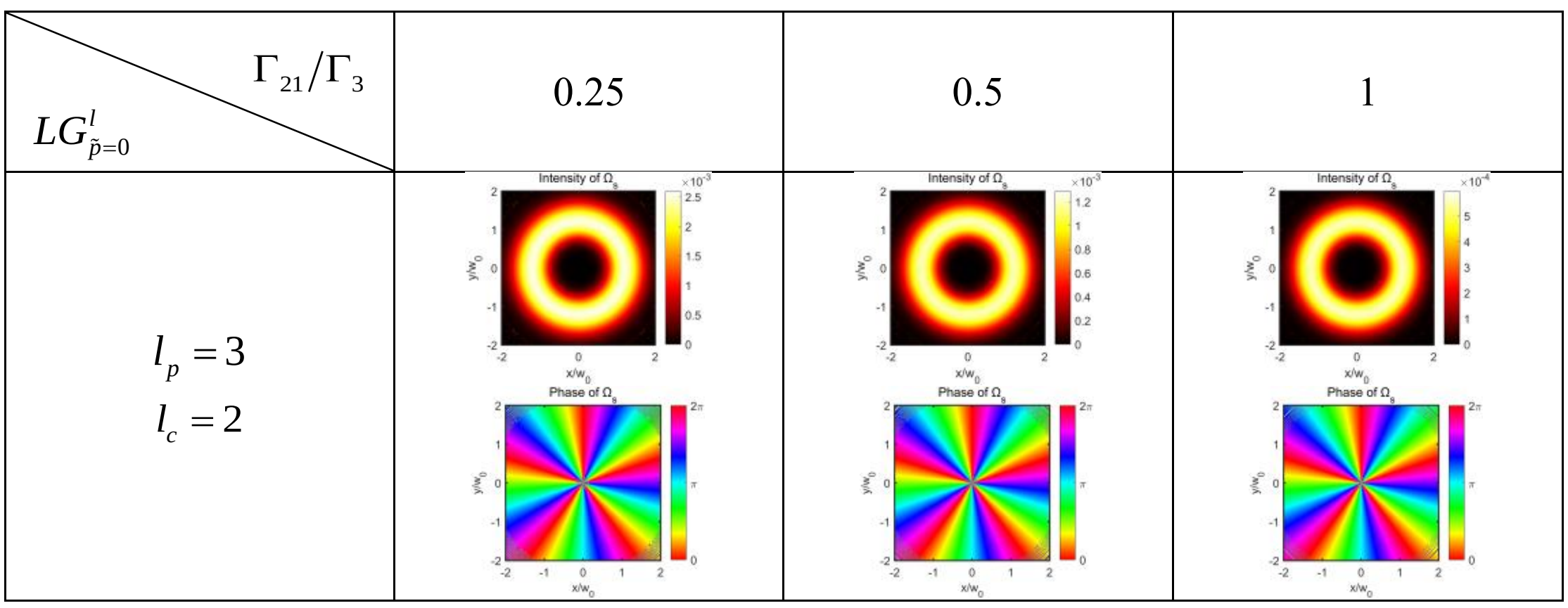

**Fig. 7(b).** During the backward propagation process, when the probe field is prepared in a LG mode $LG_0^3(\tilde{p}_p = 0, l_p = 3)$ and the control field in $LG_0^2(\tilde{p}_c = 0, l_c = 2)$, the intensity and phase distributions of the generated signal field as functions of the decay rate $\Gamma_{21}$ are shown. Other parameters are set as $\Omega_{p0} = 0.1\Gamma_3$, $\Omega_{c0} = 10\Gamma_3$, $w_{p0} = w_0$, $w_{c0} = w_0$, $\Delta_p = 0$, $\alpha = 100$.

The above analysis demonstrates that the BTWM process exhibits pronounced anti-loss characteristics, thereby significantly enhancing both the gain and the fidelity of the generated vortex signal field. This behavior arises because the backward signal field is predominantly generated at the entrance of the medium ( $z=0$ ), thereby circumventing losses associated with light propagation through the medium. This effect is also evident from Eq. (12). This three-level symmetry-broken cyclic system can be realized in superconducting artificial atoms [47-49] and molecular magnets [50]. As an example, in a three-junction flux-qubit artificial atom, the three lowest eigenstates, $|0\rangle$, $|1\rangle$, and $|2\rangle$, of the static Hamiltonian $H_0$ are chosen to construct the three-level system. When the external magnetic flux is tuned to the symmetry point ( $f=\Phi_e/\Phi_0=0.5$ ), the system potential possesses a well-defined parity, whereas the microwave driving term ( $H_1(t)=I\Phi_a(t)$ ) is odd under the parity operation. Consequently, nonzero single-photon transition matrix elements exist only between states of opposite parity. If the states $|0\rangle$ and $|2\rangle$ have the same parity, the direct single-photon transition matrix element $t_{02}=\langle 0|I\Phi_a|2\rangle$ vanishes, rendering the $|0\rangle\leftrightarrow|2\rangle$ transition forbidden, whereas the $|0\rangle\leftrightarrow|1\rangle$ and $|1\rangle\leftrightarrow|2\rangle$ transitions remain allowed. By slightly detuning the magnetic-flux bias from the symmetry point ( $f\neq 0.5$ ), the parity selection rule is broken, causing $t_{02}$ to become finite and thereby activating the previously forbidden transition. Similarly, in the spin system of the [Cr(C□S□)□]$^3$□ molecular complex, the working states are chosen from the $S=3/2$ spin multiplet with spin projections $|M_S=-3/2\rangle$ , $|M_S=-1/2\rangle$ , $|M_S=+1/2\rangle$ , and $|M_S=+3/2\rangle$ . Under the conventional magnetic-dipole selection rules, the $|M_S=+3/2\rangle\leftrightarrow|M_S=-1/2\rangle$ transition is forbidden. However, by designing molecules with a smaller axial zero-field splitting $D$ and a larger transverse anisotropy ratio $E/D$ , the transverse anisotropy term $E\left(\hat{S}_x^2-\hat{S}_y^2\right)$ strongly mixes different $M_S$ components under weak magnetic fields, such that the eigenstates are no longer pure spin-projection states but rather coherent linear combinations of multiple $M_S$ basis states. As a consequence, the conventional selection rule $\Delta M_S=\pm 1$ is effectively relaxed, and transitions that were originally forbidden acquire finite dipole transition matrix elements through state mixing, allowing coherent excitation by pulsed microwave fields. Accordingly, in this molecular complex, the states with $|M_s=+3/2\rangle$, $|M_s=+1/2\rangle$, and $|M_s=-1/2\rangle$ can be assigned as the upper, intermediate, and lower levels, respectively, thereby forming a closed-loop three-level system in which the originally forbidden transition becomes accessible. Therefore, the realization of a so-called forbidden transition does not imply enabling an intrinsically impossible process. Rather, symmetry breaking renders the first-order transition matrix element nonzero, thereby making the transition spectroscopically addressable and coherently manipulable.

## 4. Conclusion

In this paper, we investigate the transmission properties and attenuation dynamics of vortex light in a three-level symmetry-breaken system, considering both forward and backward TWM processes. We find that both schemes enable high-gain vortex light transfer, with the topological charges obeying the same algebraic relationship. During the forward process, vortex light transmission exhibits periodic oscillations, with the oscillation period modulated by relevant system parameters. Increasing the decay rate $\Gamma_{21}$ not only suppresses the generated signal intensity but also alters the transverse intensity distribution through phase modulation; however, the OAM of the vortex light remains topologically stable. In the backward process, vortex light transmission features stable attenuation of the probe field accompanied by a concomitant increase in the signal field. Although increasing $\Gamma_{21}$ attenuates the overall generated signal intensity, it does not alter the phase distribution or the transverse intensity profile. This indicates strong loss-resistance, leading to substantial enhancement of both the gain and fidelity of the signal field. Moreover, under the ATS induced by a strong control field, the influence of detuning $\Delta_p$ on gain is negligible in both schemes. Optical depth affects only the rate at which the gain approaches its maximum, without altering the ultimate peak value. These results provide a significant extension of the work reported in reference [40] and further advance the understanding of interactions between vortex light and symmetry-breaking systems. Overall, the forward process is better suited for flexible and controllable dynamic modulation of optical vortices, whereas the backward process is more advantageous for high-gain, high-fidelity, and loss-resistant vortex generation. The complementary use of these two processes is expected to provide a theoretical foundation and practical guidance for OAM-based quantum communication, quantum computing, all-optical control, and integrated nonlinear photonic devices.

**Acknowledgements**

This work was supported by the National Natural Science Foundation of China (Grant Nos. 12474353 and 12474354).

**Data availability statement**

The datasets and code generated and/or analysed during the current study are available in the Zenodo repository at https://doi.org/10.5281/zenodo.21096010.

**Disclosures**

The authors declare no conflicts of interest.

## Appendix

In this Appendix, we provide a detailed account of the theoretical calculations presented in the main text. After deriving the equations of motion for the three-level ladder-type system, given in Eq. (3), it is difficult to obtain exact solutions for the propagation of the optical fields in the medium by direct substitution into the Maxwell-Bloch equations, Eqs. (6), (7), and (11). Therefore, to obtain physically meaningful analytical solutions, we employ a perturbative approximation. Specifically, when solving the coupled equations in Eq. (3), we assume that the control field $\Omega_c$ is much stronger than the probe field $\Omega_p$ and the signal field $\Omega_s$, i.e., $\Omega_c \gg \Omega_p, \Omega_s$. This assumption implies that the control field $\Omega_c$ is continuously applied and remains independent of both time $t$ and propagation coordinate $z$. We further assume that the system is initially prepared in the ground state $|1\rangle$, such that $\rho_{11}^{(0)} = 1$, $\rho_{22}^{(0)} = \rho_{33}^{(0)} = 0$, and $\rho_{ij}^{(0)} = 0 \ (i \neq j)$.

We then expand all density-matrix elements $\rho_{ij} \ (i, j = 1,2,3)$ in Eq. (3) perturbatively as follows:

$$\rho_{ij} = \rho_{ij}^{(0)} + \Omega_p \rho_{ij}^{(1)} + \Omega_s \rho_{ij}^{(1)} + \dots . \tag{A1}$$

Under these approximations, only the following subset of equations from Eq. (3) needs to be solved:

$$\frac{\partial}{\partial t}\rho_{21} = -\left(\frac{\Gamma_{21}}{2} - i\Delta_p\right)\rho_{21} + i\Omega_p\left(\rho_{11} - \rho_{22}\right) + i\Omega_c^*\rho_{31} - i\Omega_s\rho_{23}, \quad \text{(A2)}$$

$$\frac{\partial}{\partial t}\rho_{31} = -\left[\frac{\Gamma_{31} + \Gamma_{32}}{2} - i\left(\Delta_p + \Delta_c\right)\right]\rho_{31} + i\Omega_s\left(\rho_{11} - \rho_{33}\right) + i\Omega_c\rho_{21} - i\Omega_p\rho_{32}. \quad \text{(A3)}$$

Substituting the perturbative expansion in Eq. (A1) into Eqs. (A2) and (A3), applying the assumptions above, and neglecting second-order small quantities yields a fully decoupled closed set of equations:

$$\frac{\partial}{\partial t}\rho_{21} = -\left(\frac{\Gamma_{21}}{2} - i\Delta_p\right)\rho_{21} + i\Omega_p\rho_{11} + i\Omega_c^*\rho_{31}, \quad \text{(A4)}$$

$$\frac{\partial}{\partial t}\rho_{31} = -\left[\frac{\Gamma_{31} + \Gamma_{32}}{2} - i\left(\Delta_p + \Delta_c\right)\right]\rho_{31} + i\Omega_s\rho_{11} + i\Omega_c\rho_{21}. \quad \text{(A5)}$$

Because this work focuses on the steady-state response of the system, all time-dependent derivative terms can be neglected. Solving Eqs. (A4) and (A5) then gives the steady-state analytical solutions for $\rho_{21}$ and $\rho_{31}$, corresponding to Eqs. (4) and (5), respectively. Substituting these solutions into the Maxwell-Bloch equations for the FTWM process, Eqs. (6) and (7), yields the steady-state analytical solutions for the propagation of the probe field and generated signal field in the medium, given in Eqs. (8) and (9). Similarly, substituting these solutions into the Maxwell-Bloch equations for the BTWM process, Eqs. (6) and (10), gives the corresponding steady-state analytical solutions, Eqs. (11) and (12). These analytical solutions enable us to investigate the propagation dynamics of the relevant optical fields in the medium and the transfer of OAM carried by vortex light. We note that this perturbative approach has been widely used in theoretical studies [6,7,37,38,40,41] and has been shown to provide a reasonable interpretation of experimental results [51-53].